\documentclass[12pt]{article}
\usepackage{amssymb}
\def\abstract#1{\vskip 7mm 
        \begin{center}{\large Abstract}\par \smallskip
                \begin{minipage}[c]{12cm}
                        \small #1
                \end{minipage}
        \end{center}
}
\def\title#1{\begin{center}{\Large\bf #1}\end{center}}
\def\author#1{\vskip 5mm \begin{center}{#1}\end{center}}
\def\address#1{\begin{center}{\it #1}\end{center}}

\usepackage[dvips]{graphicx}

\makeatletter
\@ifundefined{lesssim}{}{}
\@ifundefined{gtrsim}{}{}
\def\vereq#1#2{\lower3pt\vbox{\baselineskip1.5pt \lineskip1.5pt
\ialign{$\m@th#1\hfill##\hfil$\crcr#2\crcr\sim\crcr}}}
\makeatother

\begin{document}

\title{%
  Bulk Fermion Stars with New Dimensions
  \smallskip \\
  {}
}
\author{%
  Nahomi Kan\footnote{E-mail: b1834@sty.cc.yamaguchi-u.ac.jp}
  and Kiyoshi Shiraishi\footnote{E-mail: shiraish@sci.yamaguchi-u.ac.jp}
}
\address{%
   Graduate School of Science and Engineering, Yamaguchi University, \\
  Yoshida,Yamaguchi-shi,Yamaguchi 753-8512, Japan
}

\abstract{
 Many efforts have been devoted to the studies of the phenomenology 
 in particle physics with extra dimensions.
 We propose degenerate fermion stars with extra dimensions
 and study 
 what features characterized by the size of extra dimensions should appear 
 in its structure.
 We find that Kaluza-Klein excited modes arise 
 for the larger scale of extra dimensions 
 and examine the conditions on which different layers should be caused
 in the inside of the stars.
 We expound how the extra dimensions affect
 on physical quantities.
}

\section{Introduction}
 Recently, physics dealing with higher dimensions has been investigated 
 \cite{ArHame,Dienes,Anton}.
 Considering extra dimensions in Grand Unified Theories (GUTs), 
 it is suggested that forces could be unified at low energy scale.
 The first string realization of low scale gravity models was given in \cite{aadd}.\footnote{
 The possibility of large extra dimensions was originally
 discussed by Antoniadis \cite{Anton}, with the relation to string theories. }
 In this model, only gravity acts on extra dimensions,
 while the standard particles on the brane have no effects.
 On the other hand, 
 a low GUT scale has been obtained in \cite{Dienes},
 in which all the gauge particles and gravity feel extra dimensions. 

 In this paper,
 we consider that gravity and bulk matters have effects on extra dimensions
 and study how the extra spaces influence the bulk matters through the gravity.
 To investigate the influence, 
 we propose a self-gravitational system, namely star, 
 which is made of degenerate Dirac fermions with the extra dimensions.\footnote{  
 A model for higher dimensional stars was also provided by Liddle, Moorhouse and Henriques 
 \cite{Liddle}.
 They considered the neutron stars in five dimensions 
 of which the fifth one is compactified into $S^{1}$, 
 and showed that the maximum mass of the stars decreases.
 However they did not consider Kaluza-Klein excited modes 
 and put some hypotheses 
 both on the equation of state and on that of conservation, 
 so arbitrariness is left.
 }
 
 When extra dimensions are compactified,
 Kaluza-Klein (K-K) excited modes arise.
 In the present study, 
 taking account of the K-K modes, 
 we consider the fermion star with extra dimensions 
 and indicate that the excited modes affect the inside of star.
 After asking for the maximum mass and the radius 
 by use of numerical calculation,
 we will make it clear what the interior structure of star should be.
 
 We begin in Sec.~2 with five dimensional theory 
 and extend the argument into the $(4+d)$ dimensional one in Sec.~3.
 Sec.~4 describes the star in six dimensions where extra dimensions are anisotropic.
 Conclusion and discussion are summarized in Sec.~5.

\section{(4+1) dimensional Bulk Fermion Star}

\subsection{Matter}

\subsubsection{(3+1) dimensions}
  In the four dimensional theory, 
  the thermodynamic potential of a fermion gas 
  with mass $m$ and half spin is

 \begin{equation}
 \Omega_{4}=-2\frac{1}{\beta}V
        \int\frac{d^{3}\textbf{p}}{(2\pi)^{3}}
        \left[\ln
        \left(1+e^{-\beta(\sqrt{\textbf{p}^{2}+m^{2}}-\mu)}
        \right)+(\mu \leftrightarrow -\mu)
        \right] ,
\end{equation}
 where we put subscript to emphasize 
 that it stands for a four dimensional quantity.
 We will restrict our system to degenerate fermion gas 
 and take the zero temperature limit.
 Then, the thermodynamic potential becomes

\begin{equation}
 \Omega_{4}(m) \equiv
       \left\{
       \begin{array}{@{\,}ll}
       -V\frac{m^{4}}{24\pi^{2}}
       \left[\frac{\mu}{m}\sqrt{\frac{\mu^{2}}{m^{2}}-1}
       \left(2\frac{\mu^{2}}{m^{2}}-5
       \right)+3\ln\left|\frac{\mu}{m}+
       \sqrt{\frac{\mu^{2}}{m^{2}}-1}
       \right| \right]  & (m<\mu)  \\
       0  & (m\geqq\mu)  
       \end{array}  .
       \right. 
 \label{ome4} 
\end{equation}
 Thermodynamical quantities will follow from Eq.~(\ref{ome4}).

\subsubsection{(4+1) dimensions}
  We will extend our argument into the five dimensional theory.
  Here we suppose that the fifith dimension is compactified into $S^{1}$  
  with a radius $b$, $b$ is not so large.
  We impose the periodic boundary condition on a wave function 
  in the fifth dimension:

\begin{equation}
 \psi(x_5)\approx{e^{ip_{5}\cdot x_5}} ~ ,
 \label{psi5}
\end{equation}
\begin{equation}
 \psi(x_5 +2\pi{b})\sim\psi(x_5)  ~ ,
 \label{periodic}
\end{equation}
 with the fifth coordinate $x_5$ and momentum $p_{5}$ .
 Eq.(\ref{psi5}) and (\ref{periodic}) give  
\begin{equation}
 p_{5}=\frac{n}{b}  \ \ \ \mbox{($n$: integer)} ~ .
\end{equation}
 Therefore the relativistic energy involving the fifth dimension is
\begin{eqnarray}
   E_5  &=& \sqrt{{\bf p} ^2 +\left( \frac{n}{b} \right) ^2 + m^2}  \\
            &=& \sqrt{{\bf p} ^2 + M^2}  ~ ,
   \label{E5}         
\end{eqnarray}
 where
\begin{equation}
 M^2\equiv\left(\frac{n}{b}\right)^{2}+m^{2} ~ .
\end{equation}
 Using Eq.~(\ref{E5}),
 the thermodynamic potential of the fermion gas with mass $m$ and half spin 
 in the (4+1) dimensional space-time is
\begin{equation}
 \Omega_{5}=-2\frac{1}{\beta}V_{4}
        \int\frac{d^{4}\textbf{p}}{(2\pi)^{4}}
        \left[\ln
        \left(1+e^{-\beta(\sqrt{\textbf{p}^{2}+M^{2}}-\mu)}
        \right)+(\mu \leftrightarrow -\mu)
        \right] ~ .
\end{equation}
 If the integral over the fifth dimensional momentum is turned
 into the sum over quantum number $n$ :
\begin{equation}
 \int dp_{5} \to \frac{1}{b}\sum_{n} ~ ,
\end{equation}
 in addition,
\begin{equation}
 V_{4}=2\pi{b}V  ~ ,
\end{equation}
 then the thermodynamic potential is
\begin{eqnarray}
 \Omega_{5}&=&-2\frac{1}{\beta}V\sum_{n}
        \int\frac{d^{3}\textbf{p}}{(2\pi)^{3}}
        \left[\ln
        \left(1+e^{-\beta(\sqrt{\textbf{p}^{2}
        +\frac{n^{2}}{b^{2}}+m^{2}}-\mu)}
        \right)+(\mu \leftrightarrow -\mu)
        \right] \nonumber \\
        &=&
        \sum_{n}\Omega_{4}
        \left(\sqrt{m^{2}+\frac{n^2}{b^2}}
        \right)=\sum_{n}\Omega_{4}(M) ~ .
\end{eqnarray}
 (Unless $n=0$, the K-K modes become effective.)
 As well as the previous subsection, we will deal with the degenerate fermion gas 
 and low temperature near to zero.
 The thermodynamic potential, therefore, becomes
\begin{equation}
 \Omega_{5} (V,\mu,b)
         =\left\{ \begin{array}{@{\,}ll}
                           -Vb^{-4}f(\mu{b},mb)  & (\mu > M)  \\
                             0  & (\mu \leqq {M}) 
                        \end{array} ~ ,
           \right.
 \label{ome5}
\end{equation}
 where
\begin{equation}
 f\equiv\sum_{M<\mu}\frac{(Mb)^4}{24\pi^2}
 \left\{\frac{\mu{b}}{Mb}\left(
 2\left(\frac{\mu{b}}{Mb}\right)^2-5\right)
 \sqrt{\left(\frac{\mu{b}}{Mb}\right)^2-1}+3\ln\left|
 \sqrt{\left(\frac{\mu{b}}{Mb}\right)^2-1}+\frac{\mu{b}}{Mb}
 \right|
 \right\}  ~ .
\label{f51}
\end{equation}
 From Eq.~(\ref{ome5}), thermodynamical quantities are as follows:
\begin{equation}
 P=-\frac{1}{2\pi{b}}\frac{\partial\Omega_{5}}{\partial{V}}
  =-\frac{1}{2\pi{b}}\frac{\Omega_{5}}{V}=
  \frac{1}{2\pi{b}}\frac{1}{b^{4}}f  ~ ,
\end{equation}

\begin{equation}
 P_{5}=-\frac{1}{2\pi{V}}\frac{\partial{\Omega_{5}}}{\partial{b}}
 =\frac{1}{2\pi{b}}\frac{1}{b^{4}}
 \left(x\frac{\partial{f}}{\partial{x}}+y\frac{\partial{f}}{\partial{y}}
 -4f \right) ~ ,
\end{equation}

\begin{equation}
 \rho=\frac{U}{V_4}=\frac{1}{2\pi{b}}\frac{1}{b^4}\left(
 x\frac{\partial{f}}{\partial{x}}-f \right) ~ ,
\end{equation}
 where $P_{5}$ is the pressure in the fifth dimension,
 and we put ~$x=\mu{b}$ and $y=mb$.

\subsection{Space-Time}
 The line element can be written as 
\begin{equation}
 ds^2=e^{-\Phi}
 \left[-e^{-2\delta}\Delta{dt^2}+\frac{dr^2}{\Delta}+
 r^2(d\theta^{2}+\sin^{2}\theta{d\varphi^{2}})
 \right]
 +b_{0}^{2}e^{2\Phi}{d\chi}^2 ~ .
\end{equation}
 Here we regard $\Delta$, $\delta$, and $\Phi$ ($b=b_0e^{\Phi}$) 
 as functions depending only on $r$, 
 which is the distance from the origin.
 The energy-momentum tensor is
\begin{equation}
 T^{\mu}\!_{\nu}=diag.(-\rho,P,P,P,P_{5}) ~ .
\end{equation}
 We suppose isotropic pressure in the space of three dimensions
 and represent the fifth dimensional pressure as $P_{5}$.
 The equation of conservation is
\begin{equation}
 \nabla_{\mu}T^{\mu{r}}=0 ~ .
 \label{conserve5}
\end{equation}
 Eq.(\ref{conserve5}) gives the condition that the chemical potential $\mu$ satisfies
\begin{equation}
 \mu{b}=\frac{e^{\frac{3}{2}\Phi+\delta}}{\sqrt{\Delta}}\mu_{0}b_{0} ~ ,
\end{equation}
 where $\mu$ depends on $r$, and 
 $\mu_{0}$ and $b_{0}$ are the value of $\mu$ and $b$ 
 when $r$ is close to zero respectively.  
  
\subsection{Equations}
 We are just deriving the Einstein equations.
 Before that, we will rewrite Eq.~(\ref{f51}) for convenience.
 Putting
\[
 x=\mu{b} ,~~~y=mb 
\] 
 reduces Eq.~(\ref{f51}) to 
\begin{eqnarray}
 f(x,y)&=&{\sum_n}'\frac{(y^{2}+n^{2})^2}{24\pi^{2}}
 \left[
 \frac{x}{\sqrt{y^{2}+n^{2}}}\sqrt{\frac{x^{2}}{y^{2}+n^{2}}-1}
 \left(2\frac{x^{2}}{y^{2}+n^{2}}-5\right) \right.   \nonumber \\
 & & \left.
 +3\ln\left|\frac{x}{\sqrt{y^{2}+n^{2}}}+\sqrt{\frac{x^{2}}{y^{2}+n^{2}}-1}
 \right|
 \right] \nonumber \\
 &=&{\sum_n}'\tilde{f}(x,y)  \ \ \ \ \ \ \
 (\sqrt{x^{2}-y^{2}}>|n|) ~ ,
\end{eqnarray}
 where the sum over $n$ is done unless $|n|$ exceeds 
$\sqrt{x^{2}-y^{2}}$ .
 (To remark this, we put prime on a sum symbol.)
 In addition, we put
\begin{equation}
 Y=\sqrt{y^{2}+n^{2}} ~.
\end{equation}
 The leading formulae lead to
 the Einstein equations: 
\begin{equation}
 \frac{{\tilde{M}_{\star}}'}{\tilde{r}^{2}}-\frac{3}{8}
 \tilde{\Delta}(\Phi')^{2}=
 4\pi\frac{1}{m^{4}b^{4}_{0}}e^{-6\Phi}{\sum}'
 \left(x\frac{\partial\tilde{f}}{\partial{x}}-\tilde{f} \right) ~ ,
\end{equation}

\begin{equation}
 \frac{1}{\tilde{r}}\delta'+
 \frac{3}{4}(\Phi')^{2}=
 -4\pi\frac{1}{m^{4}b^{4}_{0}}\frac{e^{-6\Phi}}{\tilde{\Delta}}{\sum}'
 \left(x\frac{\partial\tilde{f}}{\partial{x}} \right) ~ ,
\end{equation}

\begin{eqnarray}
 \Phi''+\left(\frac{\tilde{\Delta}'}{\tilde{\Delta}}-\delta'+
 \frac{2}{\tilde{r}} \right)\Phi'
 &=&-\frac{8\pi}{3}\frac{1}{m^{4}b^{4}_{0}}
 \frac{e^{-6\Phi}}{\tilde{\Delta}}{\sum}'
 \left(
 \frac{3Y^{2}-2y^{2}}{Y^2}x\frac{\partial\tilde{f}}{\partial{x}}
 \right.\nonumber  \\
 & & \left.-\frac{12Y^{2}-8y^{2}}{Y^2}\tilde{f}
 \right) ~ ,
\end{eqnarray}
where
\begin{eqnarray}
 \tilde{M}_{\star} &=& \sqrt{G_{4} m^{2}} G_{4}mM_{\star} ~ , \\
 \tilde{r} &=& \sqrt{G_{4}m^{4}}r ~ , \\
 \tilde{\Delta} &=& 1-\frac{2\tilde{M}_{\star}}{\tilde{r}} ~ .
\end{eqnarray}
 In the above equations, we define a usual Newtonian constant 
$G_{4}$ as
\begin{equation}
 G_{4}\equiv\frac{G_{5}}{2\pi{b_{0}}} ~ .
\end{equation}
 Here $G_{5}$ stands for the Newtonian constant 
 in the fifth dimension.
 $M_{\star}$ is the mass interior to radius $r$.
 A prime means derivative with respect to $\tilde{r}$.
 We can solve these equations numerically.
 In the next section, we will show the result.          

\subsection{Numerical Results}
\begin{figure}
\centering
{\makebox(8,55){{\Large $\frac{M}{\bar{M}}$}}~\includegraphics[width=11cm, height=5.5cm]{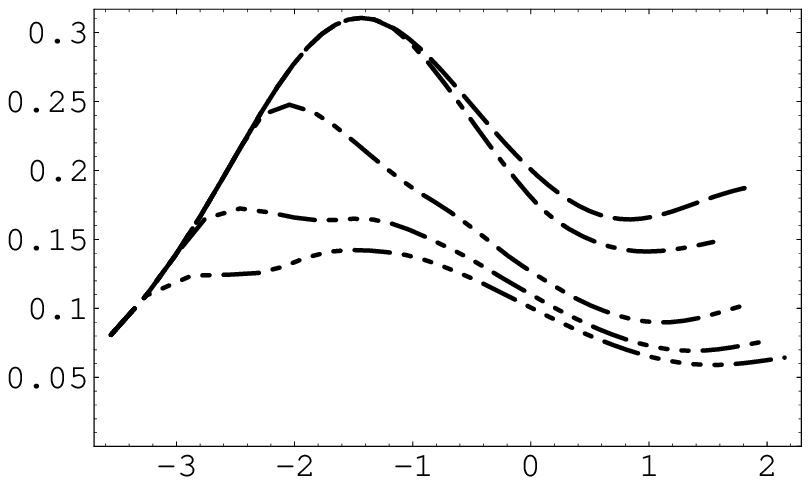}}\\
{\large $\log_{10}(\rho_{0(4)}/m^4)$}
\caption{
 Plots of the mass $M$ of the fermion stars 
 versus its central density $\rho_{0(4)}$ 
 for the various scales of the extra dimension. 
 The dashed line corresponds to $mb_0\approx 0$.
 The dot-dashed line corresponds to $mb_0=1.0$,
 the two dot-dashed line to $mb_0=2.0$,
 the three to $mb_0=3.0$,
 the four to $mb_0=4.0$.}
\label{fig51}
\end{figure}

\begin{figure}
\centering
{\makebox(8,55){{\Large $\frac{M}{\bar{M}}$}}~\includegraphics[width=11cm, height=5.5cm]{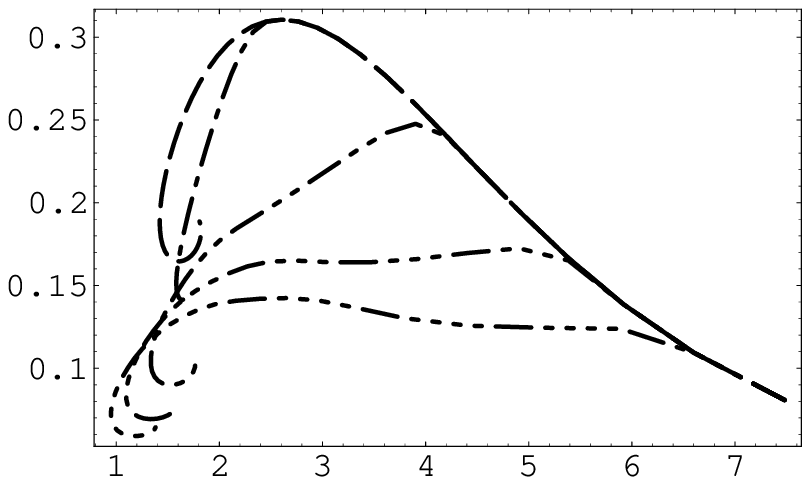}}\\
{${R}/{\bar{R}}$}
\caption{
 Plots of the mass $M$ of the fermion stars
 versus its radius $R$ 
 for the various scales of the extra dimension.
 The correspondence between the dashed lines and 
 the scales of the fifth dimension is the same as Fig.~$\ref{fig51}$.}
\label{fig52}
\end{figure}

 We will exhibit the relationship between the mass and the central density 
 in fermion stars in Fig.~$\ref{fig51}$, 
 and between the mass and the radius in Fig.~$\ref{fig52}$,
 where $\bar{M} = (G_4^{3/2} m^2)^{-1} = \frac{M_P^3}{m^2}$ 
 with the Planck mass $M_P$
 and $\bar{R} = (G_4^{1/2} m^2)^{-1}= \frac{\bar{\lambda}^2}{l_P}$
 with the Planck length $l_P$ and the Compton wave length $\bar{\lambda}$.
 These results indicate that  
 as the size of the extra space becomes larger,
 the maximum mass becomes smaller.
 Furthermore, it is remarkable that 
 two maximum points appear when $mb_{0}=3.0,4.0$.

\begin{figure}[p]
\centering
{\includegraphics[width=5cm, height=5cm]{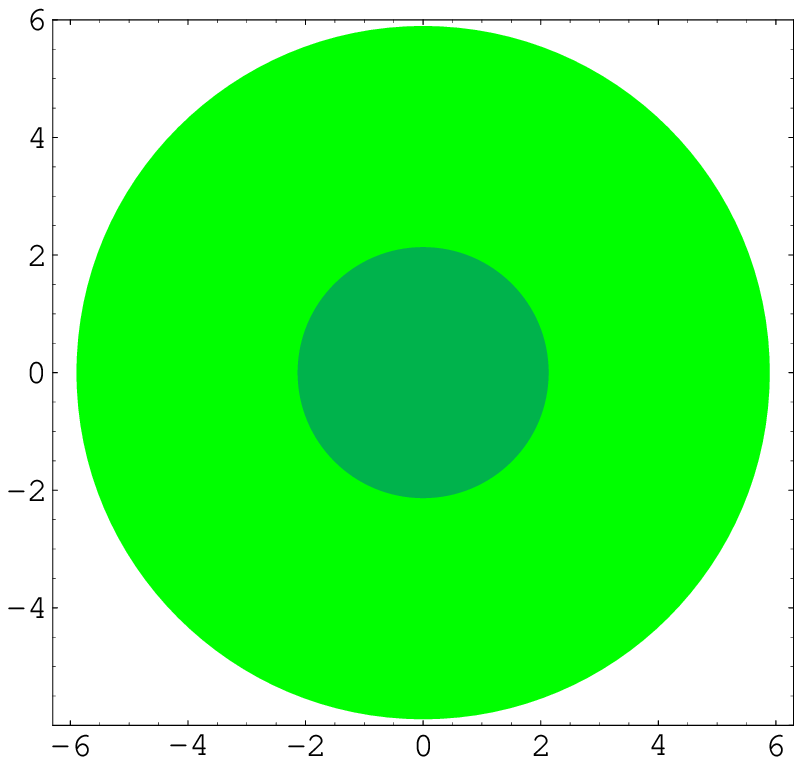}
 \hspace{0.3cm}\vspace{0.5cm}
 $mb_0=4.0$
 \includegraphics[width=5cm, height=5cm]{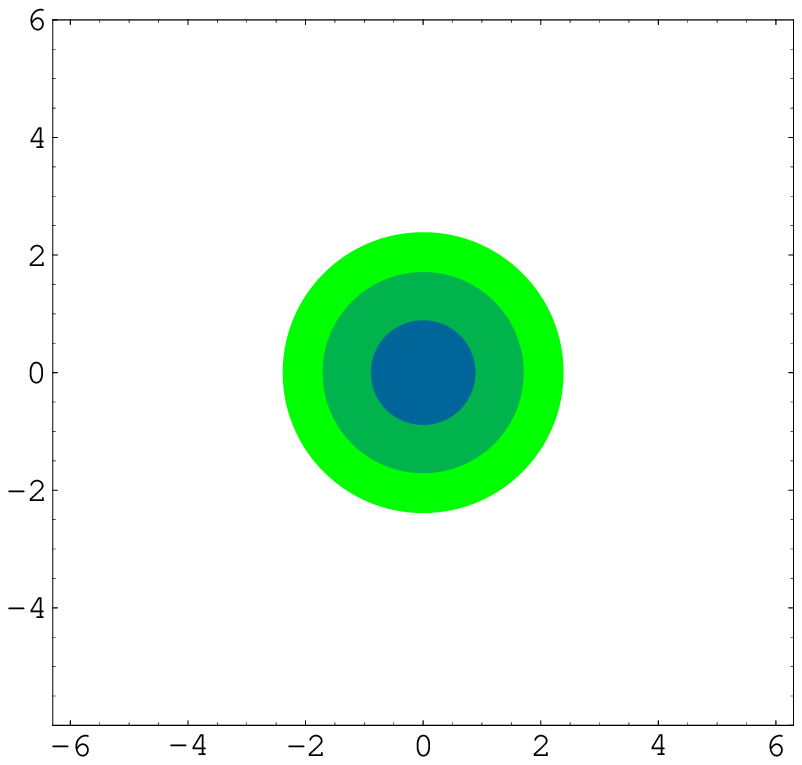}
 \hspace{1cm}\vspace{0.5cm}
 \includegraphics[width=5cm, height=5cm]{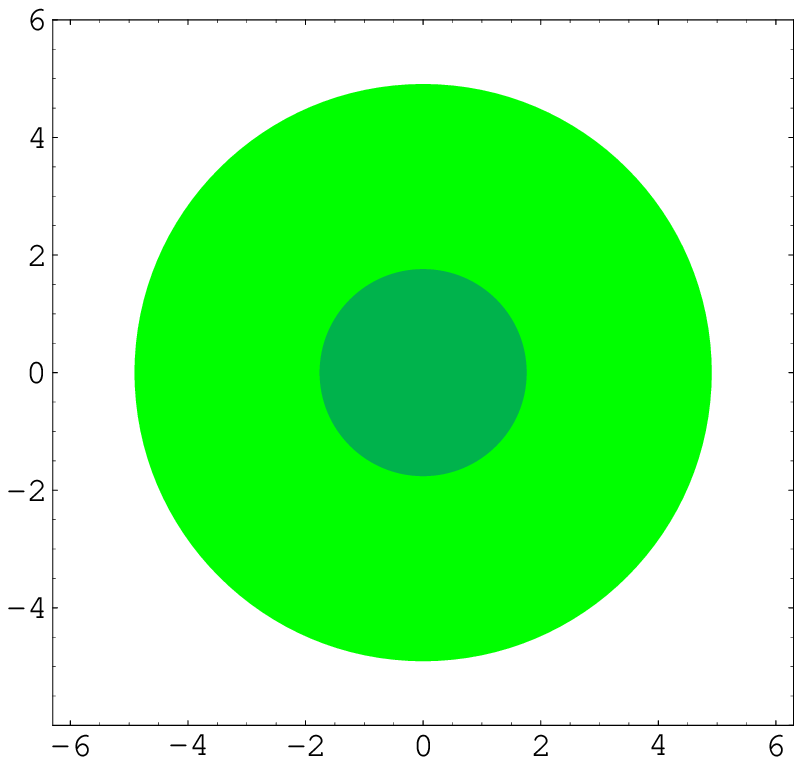}
 \hspace{0.3cm}\vspace{0.5cm}
 $mb_0=3.0$
 \includegraphics[width=5cm, height=5cm]{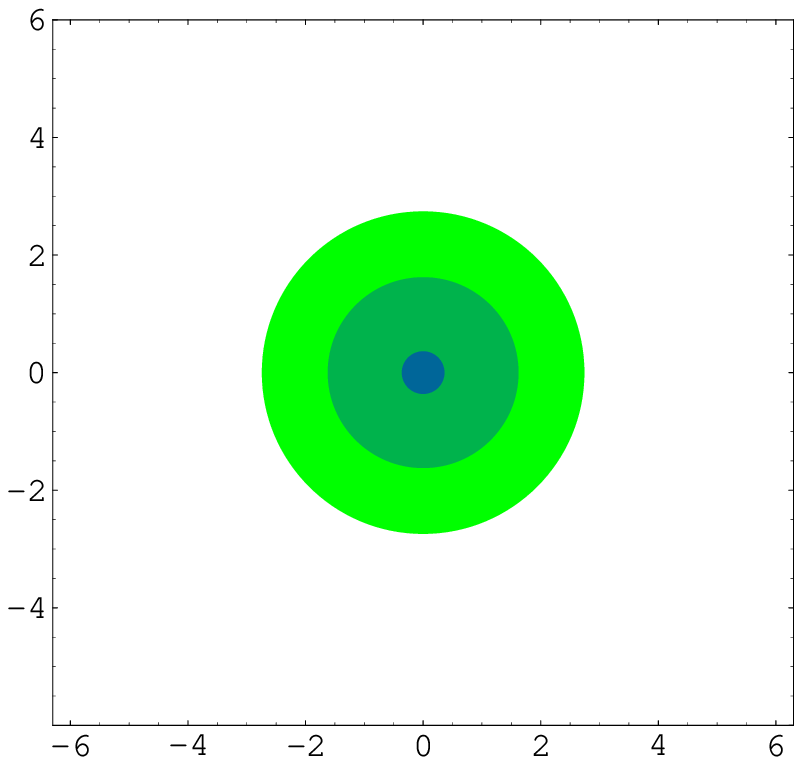}
 \hspace{12.5cm}\vspace{0.5cm}
 \hspace{1.3cm}
 \includegraphics[width=5cm, height=5cm]{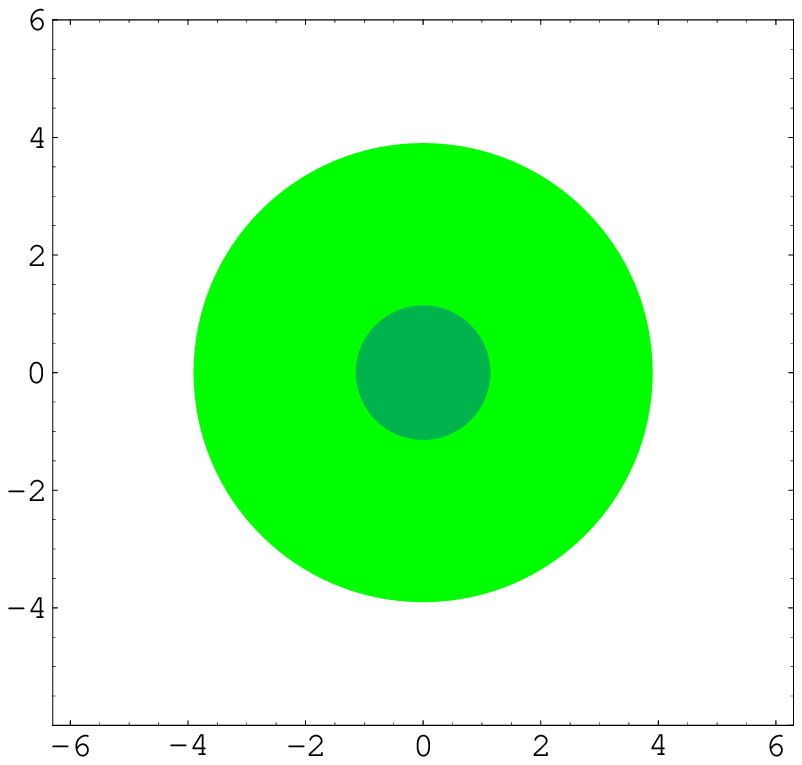}
 $mb_0=2.0$
 \hspace{12.5cm}\vspace{0.5cm}
 \hspace{1.3cm}
 \includegraphics[width=5cm, height=5cm]{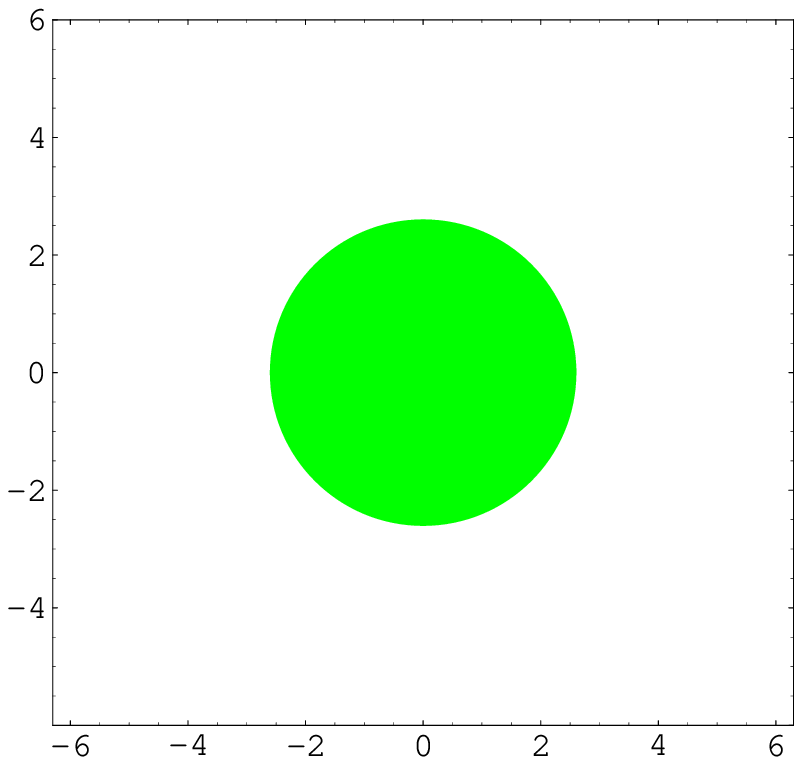}
 $mb_0=1.0$}
\caption{
 The sequence of the fermion stars with the increase of the size of the extra dimension.
 In the darkest disk, $n=2$ mode is most effective.
 In the middle one, $n=1$ mode is effective. 
 In the pale disk, there is no excited mode.}
\label{fig53}
\end{figure}

 Fig.~$\ref{fig53}$ exhibits the interior structure of the stars 
 having the maximum mass.
 From Fig.~$\ref{fig53}$,
 we find that the excited modes have effects in the core of the stars 
 as the size of the fifth dimension grows.
 We have two solutions with the maximum mass in the case of $mb_{0}=3.0,4.0$.
 For $mb_{0}=3.0$, 
 one of them is a larger star and the other is a smaller one 
 than the star with the maximum mass for $mb_{0}=2.0$ respectively.

 The central density of the larger star for $mb_{0}=3.0$ is lower 
 than that of the star for $mb_{0}=2.0$,
 while the central density of the smaller star for $mb_{0}=3.0$ is higher 
 than that of the star for $mb_{0}=2.0$,
 as Fig.~$\ref{fig51}$ shows.
 In the latter solution, a higher excited mode ($n$=2) is caused mainly 
 in the center of stars
 and a lower mode ($n$=1) in the vast region including the core.
 Proceeding to $mb_{0}=4.0$, this inclination appears more remarkably.

\section{(4+d) dimensional Bulk Fermion Stars}
 We will extend the preceding argument into the $(4+d)$ dimensional theory.
 
\subsection{Matter}

\subsubsection{(4+d) dimensions}
 We suppose that the extra dimensions are compactified into $T^{d}$  with a radius $b$, 
 $b$ is not so large.   
 Each of the momenta in the extra dimensions is  
\begin{equation}
   p_{ex}^i = \frac{n_i}{b}      \qquad  ( i =1,2, \dots , d) ~ .
   \label{pex1}
\end{equation}
 Therefore, considering the $d$ momenta, the relativistic energy is    
\begin{eqnarray}
  E_{(4+d)}  &=& \sqrt{{\bf p} ^2 +\sum_i \left( \frac{n_i}{b} \right) ^2 + m^2} \label{E4d}
        \\
                      &=& \sqrt{{\bf p} ^2 + M_{(4+d)}^2} ~ ,   
\end{eqnarray} 
 where
\begin{equation}
   M_{(4+d)} = \sqrt{\sum_i \left( \frac{n_i}{b} \right)^2 + m^2} ~ .
   \label{M4d)}
\end{equation}   
 Using Eq.~(\ref{E4d}), 
 changing the integral over the extra spatial momenta
 into the sum over quantum number $n_i$
 and holding the replacement
 $V_{(3+d)}  \to \left( 2\pi b \right) ^d V$,
 the $(4+d)$ dimensional thermodynamic potential for a fermion gas  
 with mass $m$ and half spin is
\begin{equation}
 \Omega_{(4+d)}=\frac{2^{[\frac{4+d}{2}]}}{4}
        \sum_{n_1}\sum_{n_2}\cdots\sum_{n_d}\Omega_{4}
        \left(\sqrt{m^{2}+\frac{{\bf n}^2}{b^2}}
        \right) 
        \label{ome4d}~ ,
\end{equation}
where
\begin{equation}
{\bf n}^2=\sum_{i=1}^d n_i^2 ~.
\end{equation}
(Each $n_i$ is an integer. Unless ${\bf{n}}^2=0$ , the K-K modes become effective.)
$\Omega_4$ stands for a four dimensional one.
 We will restrict our system to degenerate fermion gas 
 and take the zero temperature limit.
 Therefore,
 with the help of Eq.~(\ref{ome5}),
 the thermodynamic potential becomes
\begin{equation}
   \Omega_{(4+d)} (V,\mu,b)
           = \left\{  \begin{array}{@{\,}ll}
                                 -V{b^{-4}} f_{(4+d)} \left( \mu b, mb \right)                                  
                                        & (\mu > M_{(4+d)}) \\
                                  0    & (\mu \leqq  M_{(4+d)})
                            \end{array} ~ ,
               \right. 
               \label{ome4d2}
\end{equation}
 where 
\begin{eqnarray}
   f_{(4+d)}&\equiv& \sum \dots \sum         
                     \frac{2^{[\frac{4+d}{2}]}}{4} \frac{(M_{(4+d)}b)^4}{24\pi^2}                  
                     \left[\frac{\mu b}{M_{(4+d)}b} \sqrt{\frac{(\mu b) ^2}{(M_{(4+d)}b)^2}-1}
                              \left( 2\frac{(\mu b) ^2}{(M_{(4+d)}b)^2}-5 \right)   \right.  \nonumber \\
           & &  \qquad \qquad\qquad \qquad \qquad\quad \quad   \left.
                              +3\ln \left| \frac{\mu b}{M_{(4+d)}b}
                                                  +\sqrt{\frac{(\mu b) ^2}{(M_{(4+d)}b)^2}-1}\right| \right] ~ .
    \label{f4d}                                                 
\end{eqnarray}
The sum over $n_1,n_2, \cdots ,n_d$ are done unless $M_{(4+d)}$ exceeds $\mu$.
 From Eq.~(\ref{ome4d2}), thermodynamical quantities are as follows:
\begin{equation} 
       P = - \frac{1}{(2\pi b)^d} 
                \left( \frac{\partial \Omega_{(4+d)}}{\partial V} \right) _{\mu ,b}              
          = \frac{1}{(2\pi b)^d} \frac{1}{b^4} f_{(4+d)}   ~ ,
\end{equation}

\begin{equation} 
 P_{ex}  = -\frac{1}{(2\pi b)^d V}\frac{1}{d} b            
                    \left( \frac{\partial \Omega_{(4+d)}}{\partial b} \right) _{V,\,\mu}  
               = \frac{1}{(2\pi b)^d} \frac{1}{d} \frac{1}{b^4} 
                    \left( x\frac{\partial f_{(4+d)}}{\partial x} 
                                + y\frac{\partial f_{(4+d)}}{\partial y}- 4f_{(4+d)} \right)  ~ ,                               
\end{equation}

\begin{equation}
   \rho = \frac{U}{V_{(4+d)}}  
            = \frac{1}{{(2\pi b)}^d } \frac{1}{b^4} 
                \left( x \frac{\partial f_{(4+d)}}{\partial x} - f_{(4+d)} \right)  ~ ,               
\end{equation}
 where $P_{ex}$ is the pressure in each extra dimension,
 and we put ~$x=\mu{b}$ and $y=mb$.

\subsection{Space-Time}
 We take the line element to be of the form
\begin{equation}
 ds^2=e^{-d\Phi}
 \left[-e^{-2\delta}\Delta{dt^2}+\frac{dr^2}{\Delta}+
 r^2(d\theta^{2}+\sin^{2}\theta{d\varphi^{2}})
 \right]
 +b_{0}^{2}e^{2\Phi}\sum_{i=1}^d{d\chi_i}^2 ~ ,
\end{equation}
 where we regard $\Delta$, $\delta$, and $\Phi$ ($b=b_0e^{\Phi}$) 
 as functions depending only on $r$, 
 the distance from the origin.
 The energy-momentum tensor is
\begin{equation}
 T^{\mu}\!_{\nu}=diag.(-\rho,P,P,P,P_{ex},\cdots,P_{ex}) ~ .
\end{equation}
 We suppose isotropic pressure in the space of three dimensions
 and represent the extra dimensional pressure as $P_{ex}$.
 The equation of conservation is
\begin{equation}
 \nabla_{\mu}T^{\mu{r}}=0 ~ .
 \label{conserve4d}
\end{equation}
From Eq.~(\ref{conserve4d}), 
we find the chemical potential $\mu$ satisfies
\begin{equation}
 \mu{b}=\frac{e^{\frac{d+2}{2}\Phi+\delta}}{\sqrt{\Delta}}\mu_{0}b_{0} ~ ,
\end{equation}
where $\mu$ depends on $r$, and $\mu_{0}$ and $b_{0}$ are the values 
for $\mu$ and $b$ at $r=0$ respectively.

\subsection{Equations}
 We are just deriving the Einstein equations.
 As well as the previous section, we will rewrite Eq.~(\ref{f4d}) for convenience.
 Putting $x=\mu b$ and $y=mb$ reduces Eq.~(\ref{f4d}) to
\begin{eqnarray}
   f_{(4+d)}  \left( x, y \right)
        &=& \frac{2^{[\frac{d+4}{2}]}}{4}
                {\sum_{{\bf n}}}^{'} 
                 \frac{\left( y^2+{\bf n} ^2 \right)^2}{24\pi ^2}
                 \left[ \frac{x}{\sqrt{y^2+{\bf n}^2}} \sqrt{\frac{x^2}{y^2+{\bf n}^2}-1}
                            \left( 2\frac{x^2}{y^2+{\bf n}^2}-5 \right)  \right.     
                \nonumber \\
        &  & \left. 
                          +3\ln \left| \frac{x}{\sqrt{y^2+{\bf n}^2}}
                                               +\sqrt{\frac{x^2}{y^2+{\bf n}^2} -1} \right|                   
                            \right]     \\
        &=&{\sum_{{\bf n}}}^{'} \tilde{f}_{(4+d)} \left( x , y \right)   
           \ \ \ \ \ \ \   (x^2 - y^2  > \bf{n} ^2 ) ~ ,
     \label{f4d2}                          
\end{eqnarray}
 where 
\begin{equation}
   {\bf n}^2 = \sum_{i=1}^d {n_i}^2 ~.
\end{equation}
 The sum over ${\bf n}$ is done unless ${\bf n}^2$ exceeds 
$x^{2}-y^{2}$ .
 (To remark this, we put prime on a sum symbol.)
 In addition, we put
\begin{equation}
   Y = \sqrt{y^2 + {\bf n}^2}~.
\end{equation}
 The leading formulae lead to
 the Einstein equations:  
\begin{equation} 
  \frac{\tilde{M} _{\star} '}{\tilde{r}^2} - \frac{d(d+2)}{8} \tilde{\Delta} \left( \Phi ' \right) ^2
       = 4\pi \frac{1}{m^4 b_0^4} e^{-2(d+2)\Phi} {\sum}'
                \left( x \frac{\partial \tilde{f}_{(4+d)}}{\partial x} - \tilde{f}_{(4+d)} \right)  ~ ,               
\end{equation}         
 
\begin{equation}        
   \frac{1}{\tilde{r}} \delta ' +\frac{d(d+2)}{4} \left( \Phi ' \right) ^2 
        = -4\pi \frac{1}{m^4 b_0^4} \frac{e^{-2(d+2)\Phi}}{\tilde{\Delta}} {\sum}'
                  \left( x \frac{\partial \tilde{f}_{(4+d)}}{\partial x} \right) ~ ,                 
\end{equation}

\begin{eqnarray}
  \Phi ''  + \left( \frac{\tilde{\Delta} '}{\tilde{\Delta}} - \delta ' +\frac{2}{\tilde{r}} \right) \Phi ' 
          &=& -\frac{8\pi}{d(d+2)} \frac{1}{m^4 b_0^4} \frac{e^{-2(d+2)\Phi}}{\tilde{\Delta}} 
          \nonumber \\                 
         &&\times  {\sum}'                 
                 \left( \frac{(d+2)Y^2-2y^2}{Y^2}x\frac{\partial \tilde{f}_{(4+d)}}{\partial x} 
                            -\frac{4(d+2)Y^2-8y^2}{Y^2}\tilde{f}_{(4+d)} \right)  ~ ,                       
\end{eqnarray} 
 where 
 $\tilde{M_{\star}}(\tilde{r})=\sqrt{G_{4} m^2}{G_{4} mM_{\star}}$, 
  $\tilde{r}=\sqrt{G_{4} m^4} r$  
 and $\tilde{\Delta}=1-\frac{2\tilde{M}_\star}{\tilde{r}}$.  
 Here we define a usual Newtonian constant $G_{4}$ as 
\begin{equation}
 G_{4}\equiv\frac{G_{(4+d)}}{(2\pi{b_{0}})^d} ~ ,
\end{equation}
in which $G_{(4+d)}$ stands for the Newtonian constant 
in the $(4+d)$ dimensions.
 $M_{\star}$ is the mass interior to radius $r$.
 A prime means derivative with respect to $\tilde{r}$.
 We can solve these  numerically.
 In the next section, we will show one of the results.        

\subsection{Numerical Results}
\begin{figure}
\centering
{\makebox(8,55){{\Large $\frac{M}{\bar{M}}$}}~\includegraphics[width=11cm, height=5.5cm]{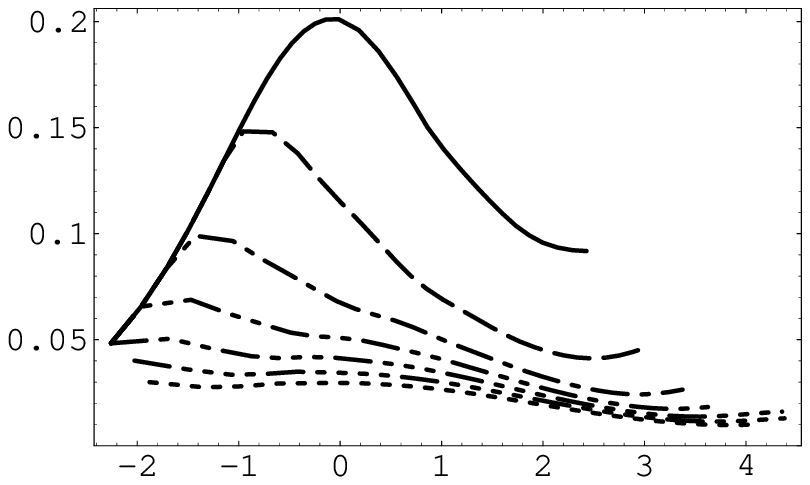}}\\
{\large $\log_{10}(\rho_{0(4)}/m^4)$}
\caption{
 Plots of the mass $M$ of the fermion stars 
 versus its central density $\rho_{0(4)}$ 
 for the various scales of the extra dimensions. 
 The solid line corresponds to $mb_0=1.0$.
 The dashed line corresponds to $mb_0=2.0$,
 the dot-dashed line to $mb_0=3.0$,
 the two dot-dashed line to $mb_0=4.0$,
 the three to $mb_0=5.0$,
 the four to $mb_0=6.0$,
 the broken line to $mb_0=7.0$.}
\label{fig61}
\end{figure}

\begin{figure}
\centering
{\makebox(8,55){{\Large $\frac{M}{\bar{M}}$}}~\includegraphics[width=11cm, height=5.5cm]{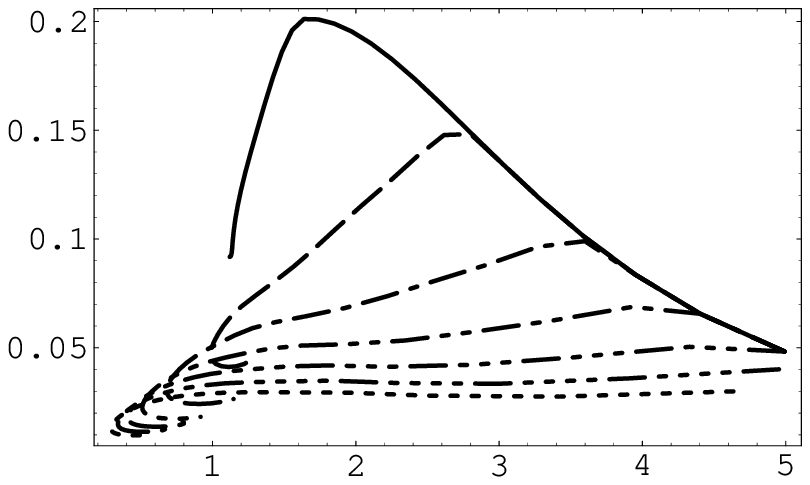}}\\
{${R}/{\bar{R}}$}
\caption{
 Plots of the mass $M$ of the fermion stars
 versus its radius $R$
 for the various scales of the extra dimensions.
 The correspondence between the lines and the scales of the extra dimensions is the same as Fig.~$\ref{fig61}$.}
\label{fig62}
\end{figure}

 We will exhibit the relationship between the mass and the central density 
 in fermion stars in Fig.~$\ref{fig61}$, 
 and between the mass and the radius in Fig.~$\ref{fig62}$ in six dimensions,
 where $\bar{M} = (G_4^{3/2} m^2)^{-1} = \frac{M_P^3}{m^2}$ 
 with the Planck mass $M_P$
 and $\bar{R} = (G_4^{1/2} m^2)^{-1}= \frac{\bar{\lambda}^2}{l_P}$
 with the Planck length $l_P$
 and the Compton wave length $\bar{\lambda}$.
 It turns out that 
 as the size of the extra space becomes larger,
 the maximum mass becomes smaller.
 Furthermore, it is remarkable 
 that two maximum points appear when $mb_{0}=5.0,6.0,7.0$.
 For $mb_{0}=5.0,6.0$, one of them is a larger star than that for $mb_{0}=4.0$
 and its central density is lower.
 In this solution, only the lowest mode (${\bf n}^2=1$) is caused
 in the center of stars.
 While the other is a smaller star and its central density is higher.
 In this solution, higher excited modes (${\bf n}^2=4,5$) are caused mainly 
 in the center of stars
 and lower modes (${\bf n}^2=1,2$) in the vast region including the core.
 Proceeding to $mb_{0}=7.0$, this inclination appears more remarkably.
 We can draw the interior structure of the stars 
 having the maximum mass, 
 but we omit them here.
 As for seven, eight and more dimensional theories, 
 the maximum mass becomes smaller with the increase of the scale of extra dimensions,
 which is similar to six one,
 but the maximum point is one and only one.
 In these solutions, the lowest mode (${\bf n}^2=1$) is caused
 in the center of stars.
 
\section{Anisotropic (4+2) dimensional Bulk Fermion Star}
 We consider the six dimensional theory.
 The  extra dimensions are compactified into $T^{2}$.
 Unlike the previous sections, 
 We suppose that 
 the scale of the compactified radius in the extra dimensions is different from each other.
 
\subsection{Matter}

\subsubsection{Anisotropic (4+2) dimensions}
 We consider that 
 the compactified radius in the fifith and sixth dimension
 is $b_1$ and $b_2$ , respectively.
 Both $b_1$ and $b_2$ are not so large.   
 Each of the momenta in the extra dimensions is  
 \begin{eqnarray}
    p_{ex}^1 &=& \frac{n_1}{b_1}  \ \ \ \mbox{($n_1$: integer)} ~,   \\
    p_{ex}^2 &=& \frac{n_2}{b_2}  \ \ \ \mbox{($n_2$: integer)} ~ .
\end{eqnarray}
Therefore the relativistic energy involving the fifth and sixth dimension is
\begin{equation}
    E_{6} = \sqrt{{\bf p}^2 +(\frac{n_1}{b_1})^2+(\frac{n_2}{b_2})^2 + m^2}  \\
       =\sqrt{\textbf{p}^{2}+M_6^{2}} ~ , 
 \label{E6}
\end{equation}
 where
\begin{equation}
 M_{6} \equiv \sqrt{\frac{{n_1}^2}{{b_1}^2}+\frac{{n_2}^2}{{b_2}^2}+m^2} ~.
\end{equation}
 Then, the thermodynamic potential for a fermion gas  with mass $m$ and half spin is
\begin{equation}
   \Omega_6       
         =2\sum_{n_1} \sum_{n_2}  
              \Omega_4 \left( \sqrt{\frac{{n_1}^2}{{b_1}^2}
                                                 +\frac{{n_2}^2}{{b_2}^2}+m^2} \right) ~.
   \label{ome61}                                                 
\end{equation}
( Unless $n_1=n_2=0$, 
 the K-K modes become effective.)
$\Omega_4$ stands for a four dimensional one.
 As well as the previous sections, 
 we will deal with the degenerate fermion gas 
 and take the zero temperature limit.
 Therefore,
 the thermodynamic potential becomes
\begin{equation}
   \Omega_{6} (V,\mu,\bar{b},\gamma)
           = \left\{  \begin{array}{@{\,}ll}
                                 -V{\bar{b}^{-4}} f_{6} \left( \mu \bar{b}, m\bar{b},\gamma \right)  
                                 =  -V{\bar{b}^{-4}} f_{6} \left( \bar{x}, \bar{y}, \gamma \right)
                                        & (\mu > M_{6}) \\
                                  0    & (\mu \leq  M_{6})
                            \end{array} ~ ,
               \right. 
               \label{ome62}
\end{equation}
 where $\bar{b}  \equiv  \sqrt{b_1 b_2},~ \gamma  \equiv  b_2/b_1,~
 \bar{x} \equiv \mu \bar{b} ,~ \bar{y} \equiv m\bar{b}$, and
\begin{eqnarray}  
   f_{6} &\equiv &
       2\sum \sum
       \frac{(M_{6}\bar{b})^4}{24\pi^2}                   
       \left[ \frac{\mu \bar{b}}{M_{6}\bar{b}}
                  \sqrt{\frac{(\mu \bar{b}) ^2}{(M_{6}\bar{b})^2}-1}
                  \left( 2\frac{(\mu \bar{b}) ^2}{(M_{6}\bar{b})^2}-5 \right)\right.  \nonumber \\
           & &  \qquad \qquad\qquad\quad \quad   \left.
                   +3\ln \left| \frac{\mu \bar{b}}{M_{6}\bar{b}}
                                        +\sqrt{\frac{(\mu \bar{b}) ^2}{(M_{6}\bar{b})^2}-1}\right| \right] ~.  
    \label{f61}
\end{eqnarray}
 The sum over $n_1,n_2$ are done unless $M_{6}$ exceeds $\mu$.
 From Eq.~(\ref{ome62}), thermodynamical quantities are as follows:
\begin{equation} 
       P = - \frac{1}{{(2\pi b_1)}{(2\pi b_2)}} 
                    \left( \frac{\partial \Omega_{6}}{\partial V} \right) _{\mu,\bar{b},\gamma}
          = \frac{1}{{(2\pi \bar{b})}^2} \frac{1}{\bar{b}^4} f_6  ~,    
\end{equation}

\begin{equation} 
       P_5 = - \frac{1}{(2\pi )(2\pi b_2) V}
                   \left( \frac{\partial \Omega_6}{\partial b_1} \right) _{V,\,\mu}   
               = \frac{1}{{(2\pi \bar{b})}^2} \frac{1}{2} \frac{1}{\bar{b}^4} 
                        \left( \bar{x}\frac{\partial f_6}{\partial \bar{x}} 
                                   + \bar{y}\frac{\partial f_6}{\partial \bar{y}}
                                    - 4f_6  -2\gamma \frac{\partial f_6}{\partial \gamma} \right) ~,                                                  
\end{equation} 

\begin{equation}         
       P_6 = - \frac{1}{(2\pi )(2\pi b_1) V}
                   \left( \frac{\partial \Omega_6}{\partial b_2} \right) _{V,\,\mu}   
               = \frac{1}{{(2\pi \bar{b})}^2} \frac{1}{2} \frac{1}{\bar{b}^4} 
                        \left( \bar{x}\frac{\partial f_6}{\partial \bar{x}} 
                                    +\bar{y}\frac{\partial f_6}{\partial \bar{y}}- 4f_6 
                                   +2\gamma \frac{\partial f_6}{\partial \gamma} \right) ~,                                                   
\end{equation}

\begin{equation}
   \rho = \frac{U}{V_6}  
            = \frac{U}{{(2\pi \bar{b})}^2 V} 
            = \frac{1}{{(2\pi \bar{b})}^2} \frac{1}{\bar{b}^4} 
                     \left( \bar{x} \frac{\partial f_6}{\partial \bar{x}} - f_6 \right) ~,                 
\end{equation}
 where $P_5$ and $P_6$ are the pressure in the fifth and sixth dimension, respectively, 
 and we put ~$\bar{x} =\mu \bar{b}$ and $\bar{y}=m\bar{b}$.
 
\subsection{Space-Time}
 The line element can be written as
\begin{equation}
   ds^2 = e^{-2\Phi} \left[ -e^{-2\delta} \Delta dt^2 + \frac{dr^2}{\Delta}
                                            + r^2 \left( d\theta ^2 + \sin ^2 \theta d\varphi ^2 \right) \right]
                + b_1^2 d\chi _1^2 + b_2^2 d\chi _2^2  ~.                                              
\end{equation}
 Here we regard $\Delta$, $\delta$, $\Phi$, $b_1$ and $b_2$
 as functions depending only on $r$, 
 the distance from the origin.
 The energy-momentum tensor is
\begin{equation}
   T^\mu\!_\nu = diag.  \left( -\rho , P, P, P, P_5, P_6 \right) ~.
\end{equation}
 We suppose isotropic pressure in the space of three dimensions
 and represent the fifth dimensional pressure as $P_{5}$
 and the sixth one as $P_{6}$.
 The equation of conservation $\nabla _\mu T^{\mu\nu} = 0$
 gives the condition that the chemical potential $\mu$ satisfies
\begin{equation}
 \mu\bar{b}=\frac{e^{2\Phi+\delta}}{\sqrt{\Delta}}\mu_{0}\bar{b}_0 ~ ,
\end{equation}
 where $\mu$ depends on $r$, and 
 $\mu_{0}$ and $\bar{b}_0$ are the value of $\mu$ and $\bar{b}$ 
 at $r=0$ respectively.  
 
\subsection{Equations}
 We are just deriving the Einstein equations.
 As well as the previous sections, we will rewrite Eq.~(\ref{f61}) for convenience.
 If we put ~$\bar{b}=\bar{b}_0 e^{\Phi (r)} , ~\gamma = b_2/b_1=e^{\phi (r)},~
 \bar{x}=\mu \bar{b}$~ and ~$\bar{y}=m\bar{b}$,~
 Eq.~(\ref{f61})  is reduced to
\begin{eqnarray}
   f_6 \left( \bar{x}, \bar{y},\gamma \right)
        &=& 2{\sum_{n_1}}^{'} {\sum_{n_2}}^{'} 
                 \frac{Y^4}{24\pi ^2}
                 \left[ \frac{\bar{x}}{Y} 
                 \sqrt{\frac{{\bar{x}}^2}{Y^2} -1}
                            \left( 2\frac{{\bar{x}}^2}{Y^2}-5 \right)  
 +3\ln \left| \frac{\bar{x}}{Y}
                                               +\sqrt{\frac{{\bar{x}}^2}{Y^2} -1}
                                                \right|                   
                            \right]     \qquad\qquad\\
        &=&{\sum_{n_1}}^{'} {\sum_{n_2}}^{'} \tilde{f_6} \left( \bar{x} , \bar{y},\gamma \right)      
           \ \ \ \ \ \ \   (\bar{x}^2 - \bar{y}^2  > {n_1} ^2 \gamma + {n_2} ^2 \gamma ^{-1}) ~ ,
     \label{f62}                          
\end{eqnarray}
 where the sums over $n_1$ and $n_2$ 
 are done unless ${n_1} ^2 \gamma + {n_2} ^2 \gamma ^{-1}$ exceeds 
$\bar{x}^2 - \bar{y}^2$ .
 (To remark this, we put prime on a sum symbol.)
 In addition, we put
\begin{equation}
   Y = \sqrt{{\bar{y}}^2 + {n_1}^2 \gamma +{n_2}^2 \gamma^{-1}}  ~. 
\end{equation}
 The leading formulae lead to
 the Einstein equations : 
\begin{equation} 
 \frac{\tilde{M} _{\star} '}{\tilde{r}^2} -\tilde{\Delta} \left( \Phi ' \right) ^2
        -\frac{1}{8}\tilde{\Delta} \left( \phi ' \right) ^2
       = 4\pi \frac{1}{m^4 {\bar{b_0}}^4} e^{-8\Phi}
                {\sum_{n_1}}^{'} {\sum_{n_2}}^{'}
                \left( \bar{x} \frac{\partial \tilde{f_6}}{\partial \bar{x}} - \tilde{f_6} \right)  ~,
\end{equation}

\begin{equation}         
  \frac{1}{\tilde{r}} \delta ' +2\left( \Phi ' \right) ^2 +\frac{1}{4} \left( \phi ' \right) ^2
        = -4\pi \frac{1}{m^4 {\bar{b_0}}^4} \frac{e^{-8\Phi}}{\tilde{\Delta}} 
                 {\sum_{n_1}}^{'} {\sum_{n_2}}^{'}
                  \left( \bar{x} \frac{\partial \tilde{f_6}}{\partial \bar{x}} \right)  ~,    
\end{equation}

\begin{eqnarray}         
  \Phi ''  + \left( \frac{\tilde{\Delta} '}{\tilde{\Delta}} - \delta ' +\frac{2}{\tilde{r}} \right) \Phi ' 
        &=& -\frac{8\pi}{8} \frac{1}{m^4 {\bar{b_0}}^4} \frac{e^{-8\Phi}}{\tilde{\Delta}} 
                  \nonumber  \\
         &&  \times
                 {\sum_{n_1}}^{'} {\sum_{n_2}}^{'}
                 \left( \frac{4Y^2-2y^2}{Y^2}\bar{x}\frac{\partial \tilde{f_6}}{\partial \bar{x}} 
                            -\frac{16Y^2-8y^2}{Y^2}\tilde{f_6} \right)   ~,         
\end{eqnarray} 

\begin{eqnarray} 
 \phi ''  + \left( \frac{\tilde{\Delta} '}{\tilde{\Delta}} - \delta ' +\frac{2}{\tilde{r}} \right) \phi '  
       &=& -\frac{8\pi}{2} \frac{1}{m^4 {\bar{b_0}}^4} \frac{e^{-8\Phi}}{\tilde{\Delta}} 
                  \nonumber  \\
         &&   \times             
                 {\sum_{n_1}}^{'} {\sum_{n_2}}^{'}
                 \left( e^{\phi} n_1^2 -  e^{-\phi} n_2^2 \right) \frac{1}{Y^2}
                 \left( 4\tilde{f_6} - \bar{x} \frac{\partial \tilde{f_6}}{\partial \bar{x}} \right) ~,                       
\end{eqnarray} 
where 
 $\tilde{M_{\star}}(\tilde{r})=\sqrt{G_{4} m^2}{G_{4} mM_{\star}}$,
 $\tilde{r}=\sqrt{G_4 m^4} r$ 
 and $\tilde{\Delta}=1-\frac{2\tilde{M}_\star}{\tilde{r}}$.  
 Here we define a usual Newtonian constant $G_{4}$ as 
\begin{equation}
    G_{4} \equiv  \frac{G_{6}}{(2\pi \bar{b}_0)^2} ~,       
\end{equation} 
in which $G_{6}$ stands for the Newtonian constant 
in the six dimensions.
 $M_{\star}$ is the mass interior to radius $r$.
 A prime means derivative with respect to $\tilde{r}$.
 We can solve these  numerically.
 In the next section, we will show one of the results.        

\subsection{Numerical Results}
\begin{figure}
\centering
{\makebox(8,55){{\Large $\frac{M}{\bar{M}}$}}~\includegraphics[width=11cm, height=5.5cm]{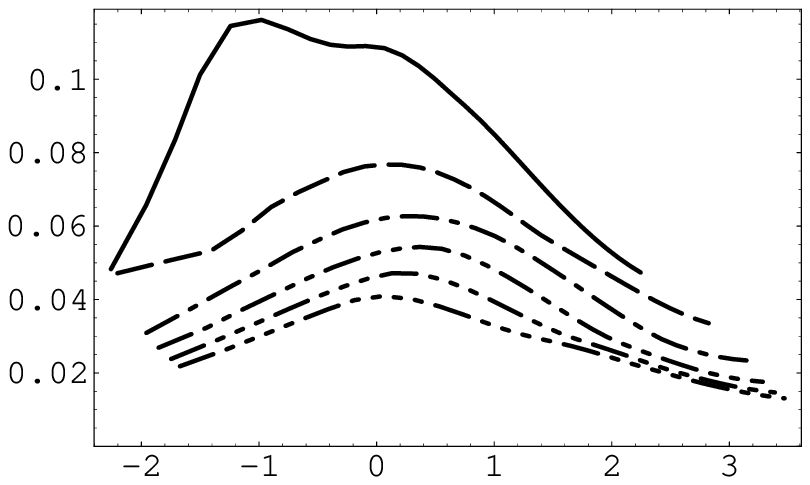}}\\
{\large $\log_{10}(\rho_{0(4)}/m^4)$}
\caption{
 Plots of the mass $M$ of the fermion stars 
 versus its central density $\rho_{0(4)}$ 
 for the various scales of the extra dimensions with $\phi _0= 2.0$. 
 The solid line corresponds to $m\bar{b}_0=1.0$.
 The dashed line corresponds to $m\bar{b}_0=2.0$,
 the dot-dashed line to $m\bar{b}_0=3.0$,
 the two dot-dashed line to $m\bar{b}_0=4.0$,
 the three to $m\bar{b}_0=5.0$,
 the four to $m\bar{b}_0=6.0$.}
\label{figID61phi2}
\end{figure}

\begin{figure}
\centering
{\makebox(8,55){{\Large $\frac{M}{\bar{M}}$}}~\includegraphics[width=11cm, height=5.5cm]{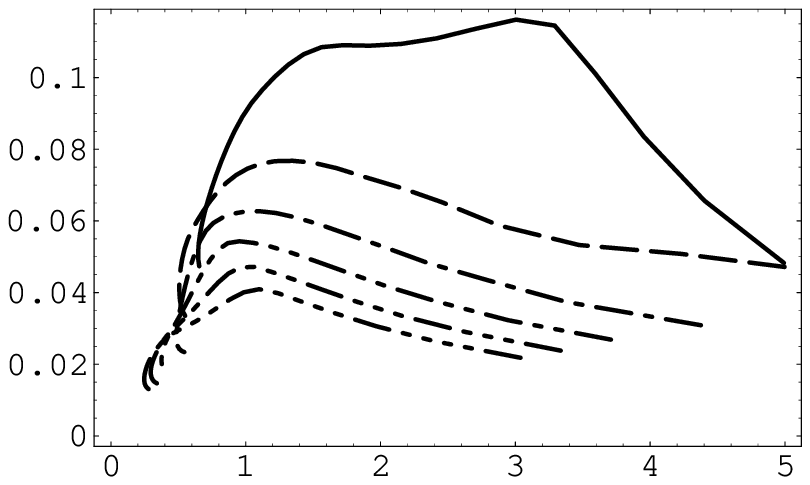}}\\
{${R}/{\bar{R}}$}
\caption{
 Plots of the mass $M$ of the fermion stars
 versus its radius $R$ 
 for the various scales of the extra dimensions with $\phi _0= 2.0$.
 The correspondence between the lines and the scales of the extra dimensions 
 is the same as Fig.~$\ref{figID61phi2}$.}
\label{figID62phi2}
\end{figure}

 We will exhibit the relationship between the mass and the central density
 in fermion stars in Fig.~$\ref{figID61phi2}$, 
 and between the mass and the radius in Fig.~$\ref{figID62phi2}$ 
 for anisotropic extra dimensions with $\phi _0= 2.0$,
 where $\phi _0$ is the value for $\phi$ 
 at the origin of the coordinates.  
 In these figures,
 we also rescale the mass and the radius by dividing 
 $\bar{M} = (G_4^{3/2} m^2)^{-1} = \frac{M_P^3}{m^2}$ 
 with the Planck mass $M_P$
 and $\bar{R} = (G_4^{1/2} m^2)^{-1}= \frac{\bar{\lambda}^2}{l_P}$
 with the Planck length $l_P$
 and the Compton wave length $\bar{\lambda}$, respectively.
 These results indicate that  
 as the size of the extra space becomes larger,
 the maximum mass becomes smaller.
 Furthermore, 
 two maximum points appear when $m\bar{b}_{0}=1.0$.
 As well as previous results,
 one of them is a larger star and its central density is lower,
 while the other is a smaller star and its central density is higher.
 In the former solution, only the lowest mode is caused 
 in the center of stars.
 In the latter solution, higher excited modes are caused mainly 
 in the center of stars
 and lower modes in the vast region including the core.
 As for $m\bar{b}_{0}=2.0,3.0,4.0,5.0,6.0$,
 the only maximum point appears. 
 These solutions are small stars with high central density and
 higher excited modes are caused in the center of stars
 and lower modes in the vast region including the core.
 We can draw the interior structure of the stars 
 having the maximum mass and 
 the relationships between the mass and the central density 
 and between the mass and the radius
 for $\phi _0= 1.0,3.0,4.0$,
 but we omit them here.            
 Instead, we will refer to the results briefly.
 For $\phi _0= 1.0$,
 two maximum points appear when $m\bar{b}_{0}=2.0,3.0$.
 One solution is the large star with low central density and
 another is the small star with high central one.
 In the former solution, only the lowest mode is caused 
 in the center of stars.
 In the latter solution, higher excited modes are caused mainly 
 in the center of stars
 and lower modes in the vast region including the core.
 When $m\bar{b}_{0}=4.0,5.0,6.0,7.0$,
 the maximum point is only one.
 These solutions are small stars with high central density
 and higher excited modes are caused in the center of stars.
 As for $\phi _0= 3.0$ and $\phi _0=4.0$, 
 we could obtain the numerical results 
 for $m\bar{b}_{0}=1.0,2.0$ and $m\bar{b}_{0}=1.0$, respectively.
 In all cases, these solutions are small stars with high central density,
 which have higher excited modes in the center of stars
 and lower modes in the vast region including the core.
 
\begin{figure}
\centering
{\makebox(8,55){{\Large $\frac{M}{\bar{M}}$}}~\includegraphics[width=11cm, height=5.5cm]{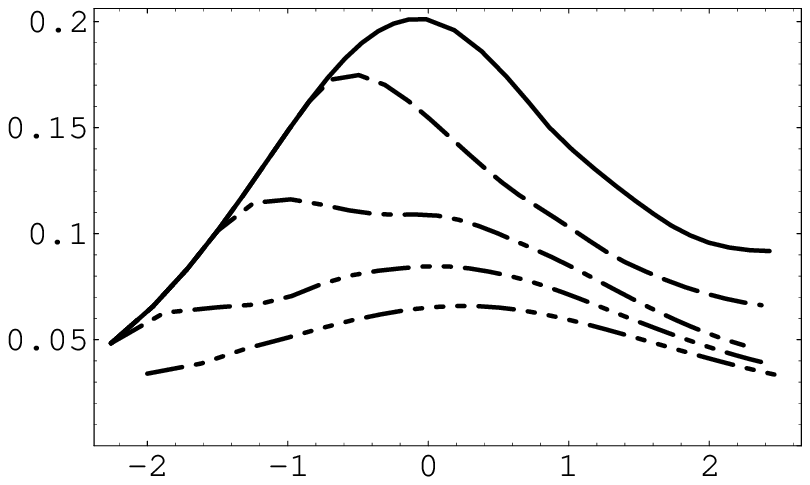}}\\
{\large $\log_{10}(\rho_{0(4)}/m^4)$}
\caption{
 Plots of the mass $M$ of the fermion stars 
 versus its central density $\rho_{0(4)}$ 
 for the various ratios of the scales of the extra dimensions with $m\bar{b}_0= 1.0$. 
 The solid line corresponds to $\phi _0=0$.
 The dashed line corresponds to $\phi _0=1.0$,
 the dot-dashed line to $\phi _0=2.0$,
 the two dot-dashed line to $\phi _0=3.0$,
 the three to $\phi _0=4.0$.}
\label{figID61mb1}
\end{figure}

\begin{figure}
\centering
{\makebox(8,55){{\Large $\frac{M}{\bar{M}}$}}~\includegraphics[width=11cm, height=5.5cm]{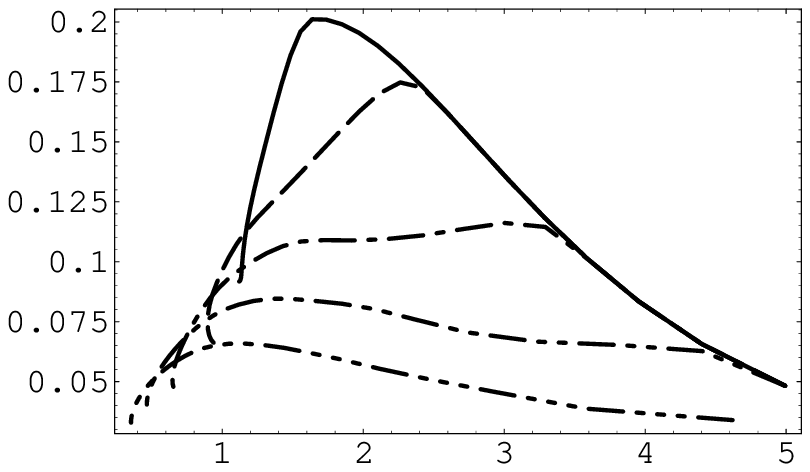}}\\
{${R}/{\bar{R}}$}
\caption{
 Plots of the mass $M$ of the fermion stars
 versus its radius $R$ 
 for the various ratios of the scales of the extra dimensions with $m\bar{b}_0= 1.0$. 
 The correspondence between the lines and the ratios of the extra dimensions 
 is the same as Fig.~$\ref{figID61mb1}$.}
\label{figID62mb1}
\end{figure}

 Fig.~$\ref{figID61mb1}$ shows the relationship 
 between the mass and the central density in fermion stars 
 and Fig.~$\ref{figID62mb1}$ between the mass and the radius  
 for anisotropic extra dimensions with $m\bar{b}_0= 1.0$.
 It turns out that 
 as the ratio of the two scales of the extra space becomes larger,
 the maximum mass becomes smaller.
 Furthermore,
 two maximum points appear when $\phi _0= 2.0$.
 As mentioned above,
 one of the solution is the large star with low central density,
 which only has the lowest mode in the center of stars, and
 the other is the small star with high central one,
 which has higher excited modes in the center of stars
 and lower modes in the vast region including the core.
 For $\phi _0= 3.0,4.0$,
 the only maximum point appears.
 These solutions are small stars with high central density
 and higher excited modes are caused in the center of stars.
 On the other hand, for the much larger scale of the extra dimensions, 
 though the anisotropy of the extra dimension is enhanced, 
 the maximum mass is almost the same.

\section{Conclusion}
 We have studied bulk fermion stars in general extra dimensional theory.
 Considering K-K modes, 
 as the scale of the extra dimensions becomes larger,
 the maximum mass more and more decreases.
 We have also obtained two sequences of solutions for five and six dimensional theories.
 One is that the fermion star is getting larger
 and its central density lower 
 with the increase of the scale of the extra dimensions,
 that is, the larger and lighter star is created.
 In these solutions, the lowest mode is only caused in the core of the stars.
 Another is that as the extra space enlarges,
 the fermion star is getting smaller and its central density higher 
 for the much larger scale of the extra dimensions, 
 namely, the smaller and denser star is created.
 In these stars,  the highest mode is caused in the center of the stars
 and the second, third and the lower modes in the vast region including the core.
 
 In the anisotropic six dimensional theory,
 as the ratio of the scales of the extra dimensions becomes larger,
 the maximum mass becomes smaller.
 We have also obtained two solutions for $\phi _0= 1.0,2.0$
 in the early stages: the scale of the extra dimensions is not so large.
 One is the large star with low central density,
 which only has the lowest mode in the center of stars, and
 another is the small star with high central one,
 which has higher excited modes in the center of stars
 and lower modes in the vast region including the core.
 However, as the scale of the extra space enlarges and  
 the anisotropy of the extra dimensions is enhanced, 
 we have obtained one solution in each investigation.
 These solutions are small stars with high central density
 and higher excited modes are caused in the center of stars.
 Especially, for the much larger scale of the extra dimensions, 
 though the anisotropy is enhanced, 
 the maximum mass is almost the same.
 
 On the basis of the above argument,
 we conclude that 
 the decrease of the maximum mass is caused by the increase of the volume of extra space
 rather than by the increase of the anisotropy of the extra dimensions.
 On the other hand, 
 the K-K excited modes become effective in the core of the stars 
 as the anisotropy of the extra dimensions is enhanced
 rather than as the scale of the extra dimensions enlarge.
 Both in the isotropic dimensional theory and in the anisotropic one,
 as the central density of the fermion stars becomes higher,
 the excited modes caused in the center of stars become higher.
 These stars tend to be getting smaller.

 As for the two sequences of solutions,   
 it is infered that the interval of the K-K excited modes
 could be related to these solutions
 from the results of the five and six dimensional theories.
 
 We have imposed no restriction on a compactified radius,
 the number of dimensions,  a unified scale {\it etc}. 
 If we set the unified scale to unity, $m \approx 1{\rm TeV}$,
 in the isotropic six dimensional theory,
 then the mass $M_{\star}$ and the radius $r_{\star}$ of the fermion star are 
 \begin{eqnarray}
   M_{\star} & \approx  & 6.5 \times 10^{23} {\rm kg}  ~,    \\
   r_{\star} & \approx  & 3.9 {\rm mm}    ~,                    
 \end{eqnarray}
 for $mb_0 =1.0$.
 Similarly, for $mb_0 =3.0$,
 \begin{eqnarray}
   M_{\star} &\approx& 3.2 \times 10^{23} {\rm kg} ~,    \\
   r_{\star} &\approx& 8.7 {\rm mm}          ~.              
\end{eqnarray}
 These stars are extraordinary small. 
 However, if we search cosmology for the evidence that the extra dimensions should be,
 this small but heavy star is candidate for unknown matter.\footnote{ 
 If we take $m \approx 1{\rm TeV}$, $mb_0 =3.0$,
  in the isotropic six dimensional theory,
 the mass of the fermion star is not excluded by the data on MACHO
 \cite{MACHO1}~\cite{MACHO2}.
 }\
 
 As far as we examine, 
 we have proved 
 that the structure of the fermion stars depends on 
 the scale of the extra dimensions, 
 that is, the excited modes have effects to the inside of stars.
 Taking the scale of the extra dimensions much larger,
 however,
 we need to analyze the stability of stars explicitly. 
 On the other hand,
 we have imposed the periodic boundary condition 
 on a wave function in the extra dimensions in this work.
 We can also adopt the general one, that is: 
$\psi(x +2\pi{b})\sim{e^{i\varphi}}\psi(x)$.
 For the anti-periodic boundary condition,
$\psi(x +2\pi{b})\sim-\psi(x)$,
 the effective mass of the fermion $M_{(h)}$ is
$M_{(h)}=\sqrt{m^2+(\frac{n\pm\frac{1}{2}}{b})^2}$.
 We have worked on the numerical calculation,
 but we have obtained no well-regulated results.
 
 As the future works, we can apply our topics to cosmology,
 in which we will suppose the time dependence of the extra dimensions
 and think over how the stars should be created in the time-dependent process.
 On the other hand,
 we took the zero temperature limit in this paper and
 we can also deal with the finite one. 
 Furthermore we will try to consider 
 what the star made of the bulk matter in the brane world should be 
 and what the star should be 
 in the extra dimensions compactified into the $d$ dimensional compact hyperbolic manifold
  \cite{CHM1}~\cite{CHM2},
  \footnote{Soliton stars with large extra dimensions are considered in \cite{SOLITON}.
  On the other hand, higher dimensional stellar solutions without compactification are studied in \cite{NONCOMPACT}.}
 in which we are going to research as the next theme. 

\section*{Acknowledgement}
 We would like to thank K.~Sakamoto and Y.~Cho for useful comments.



\begin{thebibliography}{9}
\bibitem{ArHame} N.~Arkani-Hamed, S.~Dimopoulos and G.~Dvali, 
{\it Phys.~Lett.} {\bf B429},
263 (1998); {\it Phys.~Rev.} {\bf D59}, 086004 (1999).

\bibitem{Dienes} K.~Dienes, E.~Dudas and T.~Gherghetta,
{\it Phys.~Lett.} {\bf B436},
55 (1998); {\it Nucl.~Phys.} {\bf B537}, 47 (1999).

\bibitem{Anton} I.~Antoniadis,
{\it Phys.~Lett.} {\bf B246}, 317 (1990).

\bibitem{aadd} I.~Antoniadis, N.~Arkani-Hamed, S.~Dimopoulos and G.~Dvali, 
{\it Phys.~Lett.} {\bf B436}, 257 (1998).

\bibitem{Liddle} A.~R.~Liddle, R.~G.~Moorhouse and A.~B.~Henriques, 
  {\it Class.~Quantum Grav.} {\bf 7}, 1009 (1990). 

\bibitem{MACHO1}
A.~Milsztajn and T.~Lasserre, on behalf of the EROS collaboration, astro-ph/0011375 (2000).

\bibitem{MACHO2}
E.~Kerins, astro-ph/0007137 (2000).

\bibitem{CHM1} N.~Kaloper, J.~March-Rassell, G.~D.~Starkman and M.~Trodden, 
{\it Phys.~Rev.~Lett.} {\bf 85}, 928 (2000).

\bibitem{CHM2} G.~D.~Starkman, D.~Stojkovicand and M.~Trodden, 
{\it Phys.~Rev.} {\bf D63}, 103511 (2001);
{\it Phys.~Rev.~Lett.} {\bf 87}, 231303 (2001).

\bibitem{SOLITON}
D.~Stojkovic, hep-ph/0111061 (2001).

\bibitem{NONCOMPACT} A.~Das and A.~DeBenedictis, 
{\it Prog.~Theor.~Phys.} {\bf 108}, 117 (2002).
 
\end{thebibliography}
\end{document}